# IEEE Copyright Notice

# High-Performance Spectral Element Methods on Field-Programmable Gate Arrays
## Implementation, Evaluation, and Future Projection

Martin Karp*, Artur Podobas*, Niclas Jansson*, Tobias Kenter†,
Christian Plessl†, Philipp Schlatter*, and Stefano Markidis*
*KTH Royal Institute of Technology, Stockholm, Sweden
†Paderborn University, Paderborn, Germany

*Abstract*—Improvements in computer systems have historically relied on two well-known observations: Moore's law and Dennard's scaling. Today, both these observations are ending, forcing computer users, researchers, and practitioners to abandon the general-purpose architectures' comforts in favor of emerging post-Moore systems. Among the most salient of these post-Moore systems is the Field-Programmable Gate Array (FPGA), which strikes a convenient balance between complexity and performance. In this paper, we study modern FPGAs' applicability in accelerating the Spectral Element Method (SEM) core to many computational fluid dynamics (CFD) applications. We design a custom SEM hardware accelerator operating in *double-precision* that we empirically evaluate on the latest Stratix 10 GX-series FPGAs and position its performance (and power-efficiency) against state-of-the-art systems such as ARM ThunderX2, NVIDIA Pascal/Volta/Ampere Tesla-series cards, and general-purpose manycore CPUs. Finally, we develop a performance model for our SEM-accelerator, which we use to project future FPGAs' performance and role to accelerate CFD applications, ultimately answering the question: what characteristics would a perfect FPGA for CFD applications have?

## I. INTRODUCTION

Two well-known observations have historically driven improvements in computer architectures. The first of these was observed by Gordon Moore [1], who noted that our aptitude in silicon manufacturing technologies allowed us to double the number of transistors on silicon chips every two years. Gordon Moore [1] observed the first of these. He noted that as these transistors became smaller (the transistor density increased), their supply voltage could be reduced and thus power densities remained constant (a fixed W/$mm^2$). Both of these observations facilitated the nearly five-decade improvements in computer architecture related to increasing clock frequency and instruction-level parallelism (superscalar, Out-of-order execution, etc.).

Today, Dennard's scaling has ended (since 2005 [2]), and Moore's law is literally on its last legs [3]. An initial remedy was to go towards many-core architecture (including Graphics Processor Units, GPUs [4]), albeit even those may not be enough to continue performance scaling in a post-Moore era [5]. Instead, researchers are pursuing alternative compute directions, such as disruptive neuromorphic- and quantum computing [6]. However, not all post-Moore alternatives are disruptive, with the most salient examples are reconfigurable architectures [7], such as Field-Programmable Gate Arrays (FPGAs) [8].

FPGAs aspire to be malleable and allow programmers to reconfigure the silicon to match a certain application's computational needs. Historically, FPGAs have been used to simulate Application-Specific Integrated Circuits (ASICs) before tape-out or provide hardware acceleration in (mostly fixed-point precision) low volume consumer electronics, telecommunication, or military applications. Early FPGAs were quite limited in executing floating-point operations, making them unsuitable for scientific computing. Today, this has changed, and FPGAs can execute billions of floating-point operations, rivaling existing state-of-the-art general-purpose processors (CPUs) and GPUs. More importantly, the past decade's increase in maturity of High-Level Synthesis (HLS) tools has facilitated the use of FPGAs from higher-abstraction languages such OpenCL [9], [10], C/OpenMP [11]–[13], OpenACC [14] or Java [15]. Earlier work by our peers have already demonstrated performance- or power-efficient benefits of using HLS-based FPGA designs over CPUs/GPUs in several scientific application domains [16]–[18], including Stencil-based solvers [19], [20], Molecular Dynamics [21], [22], Linear algebra [23], Neuroscience [24], [25], Network acceleration (e.g., PEACH2 [26]), and Deep Learning [27], [28]. In other application domains, FPGAs' role is still shrouded in mystery– counted among these domains is high fidelity Computational Fluid Dynamics (CFD) [29].

CFD applications are popular applications used to simulate and analyze the impact of flows on particular volumes. Fluid simulation is heavily used in aerospace analysis, biological engineering, and weather simulation, to name a few. In addition to this, CFD-solvers often occupy a large fraction of time (e.g., 40% at Beskow a Cray XC40 at KTH) and resources of supercomputers and clusters around the world, meaning that performance and power gains in CFD applications are ever so important. Spectral Element Method (SEM) based CFD solvers have ample application phases that are candidates for acceleration on FPGAs, including preconditioners and complex gather-scatter phases. However, at the core of the method, it comes down to an iterative solver evaluating the discretized system in a matrix-free fashion. The iterative solver is commonly described as the core computational component [30] in well-known high-order CFD solvers such as the incompressible flow solver Nek5000 [31]. This phase is composed of a large number of small matrix-matrix multiplications and tensor operations, which can be very vulnerable to performance degradation on modern accelerators (e.g., GPUs) as a function of problem size and characteristics [32].

In this paper, we seek to understand how FPGAs can accelerate the iterative solver in SEM-based CFD models. Also, we position our findings against existing state-of-the-art methods based on CPUs and GPUs in terms of performance (FLOP/s), power-consumption (Watts), or power-efficiency (FLOP/s/W). Finally, we derive a performance model that allows us to reason around future hypothetical FPGAs and their expected impact on these critical solvers.

We claim the following contributions:

- We design and implement an FPGA-based CFD-accelerator for the core computation **in double-precision arithmetics** of SEM-based CFD solvers using HLS that *does not sacrifice generality*,
- We empirically evaluate our proposed implementation, and position it against state-of-the-art methods using CPUs and GPUs in terms of performance (FLOP/s), power-consumption in Watts, or power-efficiency (FLOP/s/W),
- We derive an FPGA-based performance model for describing and understanding our CFD-accelerator's performance, and we use said model to project the performance (and thus role) of future FPGAs.
- We evaluate the SEM-based method with respect to performance and power-efficiency for a large number of diverse architectures, including the recently released NVIDIA Ampere-100.

## II. THE SPECTRAL ELEMENT METHOD

In this section, we provide a brief introduction to SEM and present the specific kernel for which we construct an accelerator. We have chosen to investigate a reduced problem compared to the full Navier-Stokes equations, namely the Poisson equation. In particular, we consider the local evaluation of the Poisson equation on each element, one of the core computational aspects of incompressible flow solvers such as Nek5000. The SEM solution of the homogeneous Poisson equation is based on the weak formulation, namely, calculate $u \in V \subset H_0^1$ such that,

$$\int_\Omega \nabla u \nabla v d\Omega = \int_\Omega f v d\Omega, \quad \forall v \in V \quad (1)$$

where $\Omega$ is our computational domain. The discretization is done by splitting $\Omega$ into $E$ non-overlapping elements $\Omega = \bigcup_e^E \Omega^e$ and by moving from a continuous space $V$ to a discretized space $V^N$ on a reference element with basis functions $l_i$. This choice of basis functions is unique to SEM compared to other higher order Finite Element Methods (FEM) and is based on the one dimensional $N$th order Legendre polynomials $L_N$ interpolated on the Gauss-Lobatto-Legendre (GLL) quadrature points $\xi_i$,

$$l_i(\xi) = \frac{N(1-\xi^2)L'_N(\xi)}{(N+1)(\xi-\xi_i)L_N(\xi_i)}, \quad \xi \in [-1,1].$$

The number of GLL points are related to the polynomial degree $N$ as $N + 1$. The discrete solution $u$ (in $V^N$) can then be expressed as a tensor product on the reference element,

$$u^e(\xi, \eta, \gamma) = \sum_{i,j,k}^N u_{ijk}^e l_i(\xi) l_j(\eta) l_k(\gamma)$$

where $u_{ijk}^e$ are the weights for each basis function and $\xi, \eta, \gamma$ correspond to the position within the reference element. Applying this expression to the weak formulation we arrive at the discrete bilinear $a(u, v)$ of (1),

$$a(u,v) = \sum_e^E (v^e)^T \mathbf{D}^T \mathbf{G}^e \mathbf{D} u^e = \sum_e^E (v^e)^T A^e u^e$$

where we introduce the tensor $G^e$ composed of the geometric factors that map to and from the reference element and $\mathbf{D}$ which holds the local derivatives of the operand on the GLL points. The linear form $L(v)$ of (1) is formulated similarly, and the resulting linear system is often solved for by a preconditioned Krylov subspace method.

One important point is that in higher order finite element methods, and thus also SEM, forming the local element matrix $A^e$ or the full system matrix $A$ is prohibitively expensive. Therefore, the discrete system is never explicitly constructed but rather evaluated locally per element in a matrix-free fashion inside the iterative solvers (where only the result of the tensor product is needed). This local matrix-free tensor product has been identified repeatedly as one of the most important kernels in higher order FEM [30], [32]. It has, for example, been identified in the American exascale computing project's Center for Efficient Exascale Discretizations (CEED) [33] among their bake-off kernels, Bake-off kernel 5 (BK5), which closely resembles the local Poisson operator, but also considers one more geometric factor. We focus on the pure Poisson operator, as expressed in Nekbone [34], a proxy app for the larger CFD solver Nek5000. For a more detailed description of SEM-based CFD please refer to e.g. [35].

The actual implementation of the local Poisson operator, $Ax$ or BK5 is shown in Listing 1. The most important aspect to note here is that we are essentially performing small matrix multiplications in three dimensions. Each element is of size $(N + 1)^3$, because of the number of GLL points in three dimensions, and we will refer to the points in the element as Degrees of Freedom (DOF) going forward. By measuring how many DOFs we compute, in addition to FLOPs, comparisons between different polynomial degrees can be made easier since it is an explicit measure of how quickly an element is computed.

## III. ACCELERATOR IMPLEMENTATION AND OPTIMIZATION

In this section, we present our implementation for the SEM-accelerators, and we reveal many of the optimization techniques we applied to reach high performance. Our accelerator is described using OpenCL and synthesized using the Intel OpenCL for FPGA High-Level Synthesis tool, and follow a state-of-the-art methodology for using modern FPGAs in HPC environments. Our code is open-source and available at[1] for general public use, and uses CLFORTRAN developed by CASS [36] to interface Intel OpenCL from within Fortran.

### A. Baseline

We begin our accelerator design by closely mimicking the design from Listing 1. This initial version is very similar to an earlier OpenACC-based implementation targeting GPU [37], but without the need to use large temporary arrays that cover all the elements (the input set). We can then, unlike the GPU

---
[1] https://github.com/MartinKarp/Nekbone

Listing 1: Initial code for $Ax$ (in its simplest form) for polynomial degree $N$ and $N+1$ GLL points.

```
1   void Ax(double * u, double * w, double * gxyz,
2           double * dx, double * dxt, int tot_nDOFs){
3   int nx = N+1;
4   for(unsigned ele = 0; e < tot_nDOFs; e += nx*nx*nx){
5       double shur[nx*nx*nx];
6       double shus[nx*nx*nx];
7       double shut[nx*nx*nx];
8       for(unsigned k = 0; k < nx; k++){
9       for(unsigned j = 0; j < nx; j++){
10      for(unsigned i = 0; i < nx; i++){
11          int ij = i + j*nx;
12          int ijk = ij + k*nx*nx;
13          double rtmp = 0.0, stmp = 0.0, ttmp = 0.0;
14          for(unsigned l = 0; l < nx; l++){
15              rtmp += dxt[l+i*nx] * u[l+j*nx+k*nx*nx+ele];
16              stmp += dxt[l+j*nx] * u[i+l*nx+k*nx*nx+ele];
17              ttmp += dxt[l+k*nx] * u[ij+l*nx*nx+ele];
18          }
19          shur[ijk] = gxyz[0+6*ijk+ele*6] * rtmp
20                    + gxyz[1+6*ijk+ele*6] * stmp
21                    + gxyz[2+6*ijk+ele*6] * ttmp;
22          shus[ijk] = gxyz[1+6*ijk+ele*6] * rtmp
23                    + gxyz[3+6*ijk+ele*6] * stmp
24                    + gxyz[4+6*ijk+ele*6] * ttmp;
25          shut[ijk] = gxyz[2+6*ijk+ele*6] * rtmp
26                    + gxyz[4+6*ijk+ele*6] * stmp
27                    + gxyz[5+6*ijk+ele*6] * ttmp;
28      }}}
29      for(unsigned k = 0; k < nx; k++){
30      for(unsigned j = 0; j < nx; j++){
31      for(unsigned i = 0; i < nx; i++){
32          int ij = i + j*nx;
33          int ijk = ij + k*nx*nx;
34          double wijke = 0.0;
35          for(unsigned l = 0; l < nx; l++){
36              wijke += dx[l+i*nx] * shur[l+j*nx+k*nx*nx];
37              wijke += dx[l+j*nx] * shus[i+l*nx+k*nx*nx];
38              wijke += dx[l+k*nx] * shut[i+j*nx+l*nx*nx];
39          }
40          w[ijk + ele] = wijke;
41  }}}}}
```

code, only keep one element's work arrays (the (shur, shus, and shut arrays) in on-chip memory rather than working on global work arrays for all elements.

This version, which we will call baseline, post-synthesis, validation, and execution, yielded a performance of 0.025 GFLOP/s, consumed 0.014 GB/s of external memory bandwidth, and more-than 50% of FPGA resources, which was quite discouraging. Among the reason for this initial poor performance was the low level of instruction-level parallelism, in order instructions, intermediate rounding and poor temporal locality that our accelerator had.

*B. Improving Instruction-level Parallelism and Spatial/Temporal Locality*

Our baseline solution suffers from both low locality and poor Instruction-Level Parallelism (ILP). The poor ILP indirectly leads to narrow non-coalesced accesses to the wide (up-to 2048 bit wide) on-chip Avalon memory bus. To remedy this, we extend the accelerator in two ways.

The first optimization is to improve data locality, which we perform by leveraging unique FPGA resources known as BRAMs. BRAMs are essentially small SRAM-based scratchpad memories which the user can use to store local data. Hence, by preloading and placing data with high temporal locality (gxyz, u, dx, and dxt) into these BRAMs, we significantly reduce the cost of accessing data located outside the FPGAs (in DRAM). Implementation-wise, we perform this by splitting the loop into four nested loops where we preload the data into BRAM.

We then compute on the BRAM, before ultimately writing the data back to external storage, each in separate loop nests. This optimization, by itself, already provides several benefits. Unfortunately, the access patterns into the gyxz (the geometric factors) are hard for the compiler to analyze, which results in resource *arbitration*– multiple producers/consumers have to take the turn to access the BRAM structure, leading to a large number of pipeline stalls/bubbles. To remedy and remove the arbitration, we split gyxz into six individual vectors, helping the compiler analysis pass in finding these patterns.

The second optimization concerns increasing the amount of instruction-level parallelism of the accelerators. We control the amount of instruction-level parallelism by fully unrolling selected loops (lines 14 and 35 in Listing 1) in our kernels, which effectively leads to vectorizing many instructions and providing benefits similar to SIMD-execution in general-purpose processors. Furthermore, by manually analyzing the program code, we observe that other loops can also be unrolled *only* by powers of 2 without incurring any arbitration on local BRAMs containing shur/shus/shut only if the number of GLL points, $N+1$, is divisible by the unroll factor. This observation poses a constraint in the performance of supporting GLL points that are odd, limiting how much we can unroll (through arbitration) of the accelerator (which we will see later in the evaluation). Unrolling loops also comes with the added benefit of widening the accesses to local and external memory. Therefore we also make sure to unroll the loops that load the data from external memory into the local BRAM, which makes the external memory access both wide and nicely coalesced, increasing memory throughput.

Finally, we observe that relaxing the order of floating-point operations (-ft-reassoc) as well as fusing operations before rounding (-fft-contract=fast) significantly reduce the amount of logic required for implementing double-precision Floating Points Units (FPU). Without these flags, the amount of resources required to implement double precision arithmetic on the chip quickly becomes unfeasible.

By extending our accelerator with more ILP, increased locality, and better (and more efficient) memory bandwidth, we improve the performance over the baseline by $400\times$, leading to a raw performance of 10 GFLOP/s for $N=7$.

*C. Improving Data-path Throughput*

Our accelerator is a data-flow accelerator, where data is streamed through the accelerator, and results are written back. Our accelerator is heavily pipelined, allowing the loops' live-in values to be inserted into loop body *every cycle*– an *initiation interval* of 1 cycle. Unfortunately, Intel's compiler was unable to detect this and instead generated a data-path with an *II=2* a factor 2x slower than what we calculated should be possible. This optimization is more of an observation, in which Intel OpenCL documentation discourages users from setting *II=1* (#pragma ii 1) for loops on the critical path, which we discovered is faulty, since it was *only* by using said compiler directive that the compiler finally understood that our data-path could indeed (as we designed) operate on data every cycle. The one down-side Intel warns of is a significantly reduced clock speed, but the operating frequency was virtually unchanged in our experience. This optimization (or the unexpected behavior from the compiler) yields an improved performance, reaching 60 GFLOP/s for $N = 7$.

*D. Optimizing External Memory Performance*

By default, the accelerator and Intel OpenCL runtime system will allocate data such that it is interleaved across all of the FPGAs external memory banks; in our case, data is interleaved across four banks. Typically, interleaving data increases the memory bandwidth that accelerators on the FPGA experience, but seldom can reach peak bandwidth performance [38]. Instead, for an accelerator with well-behaved and known access patterns, it is much more useful to allocate data such that each array (seen from the software perspective) is allocated on one particular bank. The reason why banking the data performs better than interleaving is that the reduced bus-arbitration between different bus masters (Intel Avalon Masters) on the accelerator is trying to access the same interleaved memory bank. In our accelerator, we split `gxyz` into six parts (see above optimization), which – together with the other data inputs – yield eight different data regions, which is split across our four memory banks. This optimization, however trivial, yields a peak performance of 109 GFLOP/s for $N = 7$.

*E. Padding*

The arbitration issue on the shared memory arrays is present when the number of GLL points, $N + 1$, is not divisible by a power of two. Therefore, we investigate padding on the host as to utilize kernels for larger degrees with a higher throughput on smaller element sizes affected by arbitration. This is successful, but padding on the host might be an issue when using the kernel in a real-world application. In addition to this, for the even number of GLL points that we focus on, the performance benefits are negligible, in particular for small elements, which we illustrate in the next section. In the end, we do not use padding for our fully synthesized kernels.

## IV. PERFORMANCE MODEL

In this section, we introduce a performance model for our accelerator. The model takes into account the bandwidth and the computational aspects of our accelerator. It also considers FPGA resources, which are broken down into three types: **(i)** Digital Signal Blocks (DSP), which dictates the number of multiplications the FPGA can perform, **(ii)** Programmable Logic, which is where all most functionality is encoded into, and **(iii)** BRAM, which are SRAM-based scratchpad memories used to store data on the FPGA. The goal is to provide a model general enough to project the performance and resource utilization for current and future FPGAs and memory systems. First, we introduce a cost measure for the kernel

$$C(N) = (\text{adds}(N), \text{mults}(N)) = (6(N+1)+6, 6(N+1)+9)$$

where we take into account the number of Adds and Mults that are executed per DOF in an element. This is only a function of the polynomial degree $N$ and independent of the number of elements. In addition to this we are interested in the number of data accesses to global memory which per DOF is

$$Q(N) = (\text{loads}(N), \text{writes}(N)) = (7, 1).$$

The single write and six reads per DOF can be attributed to the accesses to `w` and `gxyz` respectively in Listing 1, whereas the seventh read comes from the accesses to `u` after exploiting all reuse potential of `u` values within each element.

With these two measures we have the operational intensity

$$I(N) = \frac{12(N+1)+15}{8 \cdot S}, \quad S = \texttt{sizeof(double)}$$

enabling us to relate the bandwidth to the performance with the roofline model [39]. For a conventional processor these measures would maybe be enough for us to relate the performance to bandwidth and how many operations that can be executed per cycle. However, the roofline model is a quite crude tool that does not take into account the requirements for the implementation or the limitations of our hardware. We now seek to make a more detailed performance model that take these aspects into account.

Since we use re-configurable hardware, the resource utilization for our kernel and its relation to the cost and bandwidth is another important measure. Therefore, we introduce the resource measure related to the amount of Digital Signal Processors (DSP), logic in the form of Adaptable Logic Modules (ALM), as well as the amount of shared memory in the form of BRAM we need for our computation as

$$R_{\text{tot}} = (\#\text{DSPs}, \#\text{ALMs}, \#\text{BRAM}) = R_{\text{base}}(N) + R_{\text{comp}}(N).$$

This measure is composed both by the base resource utilization $R_{\text{base}}$, that can be empirically measured for each degree and is independent of the throughput and the number of adds and mults we need to support per cycle. The resource utilization caused by the compute, $R_{\text{comp}}$, can be estimated by introducing $R_{\text{add}}$, $R_{\text{mult}}$. These constants are the number of DSPs and ALMs necessary to implement a multiplication or an add on our FPGA. By empirically measuring these, we can then relate the resource utilization to the throughput, $T$, of computed nodal points per cycle (DOFs/cycle) as the following

$$R_{\text{comp}}(N) = T \cdot (C_{\text{add}}(N) \cdot R_{\text{add}} + C_{\text{mult}}(N) \cdot R_{\text{mult}})$$

The question we then need to answer is what throughput $T_{max}$ can be achieved, and what is necessary? There is no need to have a higher throughput for our computation than the number of nodal points we can load from global memory. We can therefore express the maximum throughput based on the bandwidth, $T_B$, as a function of the bandwidth $B$ and amount of data that is necessary per DOF

$$T_B = \frac{B}{8 \cdot S}.$$

Then, depending on whether the amount of re-configurable resources on a specific FPGA, $R_{\max}(N) = R_{\text{tot}} - R_{\text{base}}(N)$, is the limiting factor or the memory bandwidth, we have that the maximum throughput per cycle can be expressed as

$$T_{\max}(N, B, R_{\text{tot}}) = \min\left\{\frac{R_{\max}(N)}{R_u}, T_B\right\}$$

where we have an element wise division over $R_{\max}$. This can then directly be related to theoretical peak performance, $P_{\max}(N)$, as

$$P_{\max}(N) = (12(N+1) + 15) \cdot T_{\max} \cdot f$$

by taking the cost, $C(N)$, and operating frequency, $f$, of the FPGA/global memory system into consideration.

At this point, our model is oblivious to any other constraints than the bandwidth, resources available and resources necessary to perform our operations, and provides a clear estimation of the performance of our kernel. The model is also general enough to project performance for Xilinix FPGAs. However, to nuance the model, and match achieved throughput, we also need to take other limitations when designing our accelerator into account. As we discussed in the previous section, the effects of arbitration are a limiting factor for simulation performance with certain polynomial degrees. This observation leads us to introduce further constraints on $T$, depending on the element dimensions, which can be expressed as

$$T \in 2^k, \quad k \in \mathbb{Z}_0^+$$
$$N + 1 \mod T = 0.$$

The second constraint could be avoided by padding, but the execution time and resource utilization would then simply reduce to that the closest polynomial $N_2$ such that we can increase the throughput to $T_2$. The effective amount of compute necessary would then increase and the performance gain by padding $p$ can be expressed as

$$T = T_2 \left( \frac{N+1}{N+1+p} \right)^3$$

where $T_2$ would be the new vector length and $p$ the amount of padding. From this expression, we can see that for most degrees, in particular small ones, padding would simply decrease the performance. However, we note that the need for padding and the cause of arbitration is an issue that might be addressed in future HLS software. Therefore, for the projections of the model we focus on polynomial degrees which enable us to efficiently control the throughput/unroll without arbitration.

However, throughput is not the only limiting factor on an FPGA. We must also consider the amount of memory, block RAM, available on the FGPA as well. For some FPGAs, the BRAM size might be a limiting factor rather than the number of DSPs and logic. However, compared to the throughput $T$, the required memory, $R_{\text{mem}}$, depends only on the element size and, therefore, is independent on different platforms. This means that we can use experimental results from one platform to project the BRAM utilization for future FPGAs.

## V. RESULTS

In this section, we first go through our experimental hardware, software, and our experimental methodology. After this, we present our measurements for our SEM-accelerator and compare it to other computing platforms. Finally, we discuss the future outlook for FPGAs as an accelerator for the family of higher order Finite element Methods, with a focus on SEM.

### A. Experimental Platform

Intel OpenCL SDK for FPGAs [9] version 20.2 (Quartus Prime v19.4) was used to implement and design our accelerator. The Bittware 520N platform [2] on the Paderborn Noctua FPGA cluster [18], which contains a Stratix 10 GX2800 device and four banks of DDR4 memory, was used to evaluate our SEM-accelerator. We compare and contrast the performance of our accelerator against three state-of-the-art CPUs and five different generations of NVIDIA GPUs (including the recent Ampere-series). An overview of all architectures that we used can be seen in Table II, which also includes their relative strengths and weaknesses.

For the CPU version, we use the state-of-the-art implementation of $Ax$ found in Nekbone [34], compiled with the standard optimization flags and recent compiler versions, and executed the kernel with one MPI thread per CPU core. For the GPU version, we use the optimized version described in [40], which has been shown to reach close to the implementation-derived roofline's peak performance for $N = 9$. Power-consumption measurements are performed using Intel RAPL counters [3] and Marvell tx2mon kernel module [4] for the CPUs, NVIDIA Management Library (NVML) [5] for the GPUs, and Bittware provided MMD functions for the FPGA board that can be accessed through an API in OpenCL. We use the largest data input set for the power-consumption measurement to get stable readings. All experiments are executed to *exclude* PCIe transfer overheads, focusing exclusively on the isolated performance of the kernel. We omit error bars in our graphs since the standard deviation is smaller than 2%.

### B. SEM Accelerator Synthesis and Performance Summary

We created eight SEM-accelerators, where each discrete accelerator is specialized to support a different polynomial degree $N$. The resource utilization and performance characteristics of each of these accelerators can be seen in Table I. The operating frequency of our accelerators ranges between 170 and 391 MHz. Overall – assuming that the accelerator is driven by external memory – we usually expect a linear performance increase as a function of the frequency of to 300 MHz, after which the performance no longer improves. This is due to the memory controllers, which operate at a 300 MHz frequency (@512-bit/cycle throughput). Except for one corner case $N = 13$, all of our accelerators reach 200 MHz (and above) clock frequency, which allows our accelerators to exploit close to the peak available bandwidth of our system. Among the three different types of resources that an FPGA provides (Logic, DSPs, and BRAM), our accelerator consumes and is bound primarily by the amount of logic on the device. This observation is quite unique, as previous FPGA accelerators for HPC use have been primarily bound by the number of DSP blocks, the number of BRAM blocks, or both (See e.g., [19] for an example involving Stencils). Our accelerator is bound by logic primarily due to us using double-precision arithmetic, which still is a non-negotiable requirement in higher order FEM [6]. In summary, the FPGAs produced today are oversized with respect to both BRAM and DSPs for CFD-like computation, and the silicon area these occupy should ideally be invested into more logic. Our accelerators' performance varies with the polynomial degree, where the accelerators supporting $N = 7, 11, 15$ reach the highest performance. The performance we achieve for degrees seven and 11 is especially important since most simulations utilize a polynomial base in this range [41]. The peak performance was measured to be as high as 109, 136.4, and 211.3 GFLOP/s, respectively, favoring higher polynomial degrees since the operational intensity increases with higher order. Lower polynomial orders experience a rather low performance, reaching as low as 22.1 GFLOP/s for the lowest degree, which can be attributed to the low operational intensity, but also the relatively low effective bandwidth for small input sizes. If we adjust for the frequency of the kernels and look at polynomials 7, 11, and 15, the number of DOF/cycle as shown in Table I is fairly constant (3.58, 3.96, and 3.83, respectively). This throughput is in agreement with our performance model which for this FPGA gives $T_{\max} = 4$.

---
[2] https://www.bittware.com/fpga/520n/
[3] https://www.intel.com/content/dam/www/public/us/en/documents/manuals/64-ia-32-architectures-software-developer-vol-3b-part-2-manual.pdf
[4] https://github.com/Marvell-SPBU/tx2mon
[5] https://developer.nvidia.com/nvidia-management-library-nvml
[6] Experiments with single-precision or lower may work for some scenarios, but for longer simulations in particular the cumulative error can lead to highly innacurate results

| N | $f_{\max}$ (MHz) | Logic Util. (%) | Registers | BRAMs (%) | DSPs (%) | Power (Watt) | Empirical Performance | | | Model Error (%) |
|---|---|---|---|---|---|---|---|---|---|---|
| | | | | | | | GFLOP/s | GFLOP/s/W | DOFs/cycle | |
| 1 | **391** | 31% | 539409 | 4% | 6% | 81.05 | **22.1** | **0.27** | 1.45 | **27.61** |
| 3 | 292 | 50% | 1031880 | 9% | 14% | 84.38 | 62.2 | 0.78 | 3.28 | 17.99 |
| 5 | 243 | 46% | 968793 | 10% | 5% | 77.52 | 31.4 | 0.41 | 1.48 | 25.89 |
| 7 | 274 | 72% | 1464437 | 18% | 24% | 90.38 | 109.0 | 1.21 | 3.58 | 10.05 |
| 9 | 233 | 59% | 1350551 | 27% | 7% | 84.31 | 62.4 | 0.74 | 1.98 | 0.82 |
| 11 | 216 | 69% | 1511613 | 34% | 17% | 90.65 | 136.4 | 1.50 | **3.96** | 1.02 |
| 13 | **170** | 70% | 1644011 | 53% | 10% | 83.37 | 62.14 | 0.74 | 1.99 | **0.31** |
| 15 | 266 | 77% | 1705581 | 39% | 22% | 99.65 | **211.3** | **2.12** | 3.83 | 4.30 |

TABLE I: Synthesis and performance of our SEM-accelerator when synthesized for different polynomial degrees, highlighting **highest** and **lowest** obtained metrics respectively. Notice that our SEM-accelerator is bound by the amount of logic available on the FPGA, subject primarily through the necessity of using IEEE-754 double-precision arithmetic (rather than single-precision arithmetic, as is often the case when using FPGAs).

| Type | Architecture | Tech.(nm) | Peak Perf. (GFLOP/s) | Mem. B/W (GB/s) | TDP (W) | Byte/FLOP | Freq. (MHz) | Release |
|---|---|---|---|---|---|---|---|---|
| FPGA | Stratix GX 2800 | 14 | 500* | **76.8** | 225 | 0.154 | **400** | 2016 |
| CPUs | Intel Xeon Gold 6130 | 14 | 1075 | 128 | **125** | 0.12 | 2100 | 2017 |
| | Intel i9-10920X | 14 | 921 | **76.8** | 165 | **0.083** | **3500** | 2019 |
| | Marvell ThunderX2 | 16 | 512 | 170 | 180 | 0.33 | 2000 | 2018 |
| GPUs | NVIDIA Tesla K80 | 28 | 1371 | 240 | **300** | 0.17 | 562 | 2014 |
| | NVIDIA Tesla P100 SXM2 | 16 | 5304 | 732.2 | **300** | 0.14 | 1328 | 2016 |
| | NVIDIA RTX 2060 Super | 12 | **224.4** | 448 | 175 | **2** | 1470 | 2019 |
| | NVIDIA Tesla V100 PCIe | 12 | 7066 | 897 | 250 | 0.12 | 1245 | 2017 |
| | NVIDIA A100 PCIe | 7 | **9746** | **1555** | 250 | 0.16 | 765 | 2020 |

TABLE II: Overview of selected systems used in the evaluation of the SEM-computation, including derived metrics to contrast the different architectures, highlighting the **highest** and the **lowest** observable metrics. Notice how for double-precision, the FPGA is on the lower end of the spectrum, with ample future opportunities in terms of peak performance, memory bandwidth, and base frequency. *Peak Perf. is an optimistic bound based on our performance model at 400 MHz and empirically measured resource utilization.

As for other degrees, we note an evident performance degradation when the kernel is not divisible by 4. This implies that the peak throughput in our case, and what we use in our performance model is exactly the largest power of 2 such that the number of GLL points, $N+1$ are divisible by this. However, our analysis is that this is primarily caused by limitations in the HLS and might not hold going forwards. Therefore, we focus on polynomials that enable us to unroll without arbitration for our future projections, i.e., 7, 11 and 15.

Our accelerators' power consumption varies between 80.0 and 99.65 Watts, which is a function of the device's resource utilization and frequency. If we couple the power consumption and raw performance, our accelerators can reach 1.21, 1.5, and 2.12 GFLOP/s/W of performance, which we will see (see section V-C) is better than all examined state-of-the-art CPUs.

Finally, we investigate the model error, which is the difference between the throughput per cycle predicted by our performance model (previous section IV) and what was empirically measured on the device. One thing to note is that there is a significant dependence on the problem size and the effective bandwidth. We observed this empirically and also by investigating the performance of the STREAM benchmark for FPGAs [42] and performing our measurements. Because of the input dependent bandwidth, the performance for smaller degrees is subject to a lower bandwidth than what we used in our model. This is evident as the error decreases as the polynomial degree increases. From polynomial degree seven and onwards though our model differs at most 10.05% compared to the throughput of what we empirically measure. In particular, the performance modeled for 10 and 14, where the throughput is only limited by arbitration and not the bandwidth, the error is less than 1%. The modeled performance is also shown in Fig. 3, where we can see that the modeled performance correlates very accurately with what we measure, only that the frequency fluctuates. We note that for the degrees where we are not limited by arbitration we can achieve performance very close to the roofline.

### C. SEM Accelerator Quantitative Performance Comparison

In this section, we compare the performance quantitatively of our SEM-accelerator with those achieved by CPUs and GPUs for the same input size. Fig. 1 overviews the results, where the measured performance (GFLOP/s) is shown as a function for problem size (the number of elements).

For small elements (Fig. 1:a-c), none of the systems reach very high-performance since the arithmetic intensity is low computation as well as the fact that the effective bandwidth for small inputs is in general quite low due to latency and kernel overhead; given that the FPGA that our accelerator uses is operating at the lowest frequency and lowest bandwidth (out of all the systems, see Table II), this leads to a struggle for performance of our SEM-Accelerator compared even to the CPUs for all these graphs. As $N$ increases, we note a clear distinction between the CPU and GPU systems in favor of the latter; in particular, the Tesla-class GPUs (Pascal-100, Volta-

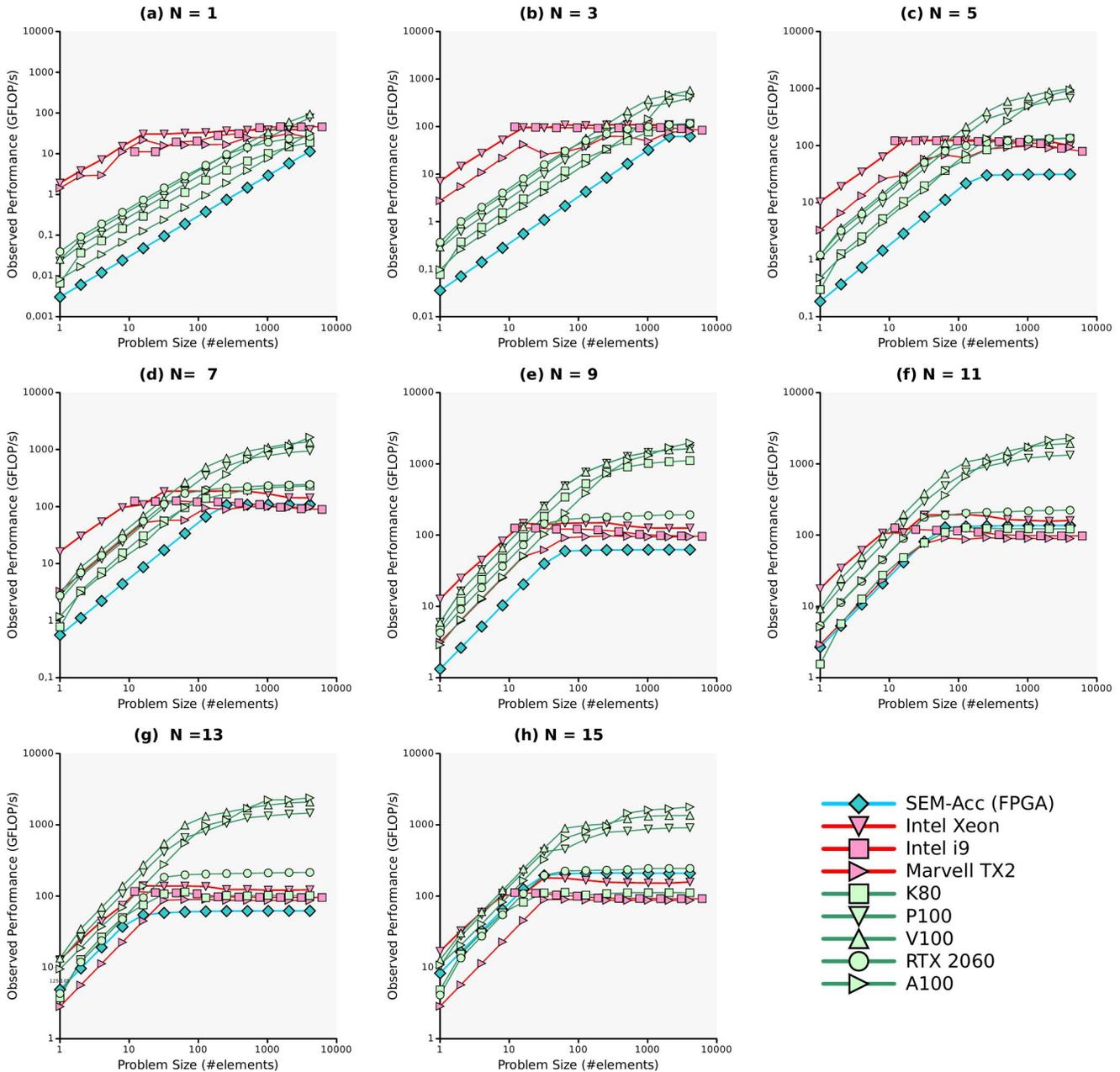

Fig. 1: Observed performance (GFLOP/s, y-axis) of the SEM-solver as a function of both problem size (# elements, x-axis) and the polynomial degree $N$, going from $N = 1$ to $N = 15$, each showcased in a separate plot. We show the performance for each $N$ for our FPGA-based SEM-accelerator (blue), state-of-the-art CPUs in red and GPUs in green.

100, and Ampere-100) distinguish themselves performance-wise from the other systems because of their high bandwidth.

For medium-sized elements (Fig. 1:d-f), we see an increase in performance for our SEM-accelerator only when using both $N = 7$ and 11; at these sizes, our accelerator outperforms both the Intel i9-10920X and the Marvell ThunderX2 processor by up-to 1.08x and 1.48x, respectively, and also outperform the Tesla-class K80 GPU by 1.07x. The reason why degree 9 underperforms on our SEM-accelerator is that we are limited in order to avoid arbitration in how much we can unroll and thus what throughput we can achieve for $N = 9$. The NVIDIA Pascal-100, Volta-100, and Ampere-100 reach 1.3 TFLOP/s, 1.9 TFLOP/s, and 2.3 TFLOP/s, respectively, surpassing all other architectures by a magnitude in performance.

For large elements (Fig. 1:d-e), only degree 13 performs poorly (with explanation above) on our SEM-Accelerator. For $N = 15$, the SEM-Accelerator reaches peak performance of 211.3 GFLOP/s, beating the Intel Xeon 6130, Intel i9-10920X, and Marvell ThunderX2 by 1.17x, 1.89x, and 2.34x, respectively. The SEM-Accelerator also outperforms both the Kepler-class NVIDIA K80 by a factor 1.87x, while having 0.86x the performance of the Turing-class RTX 2060 GPU. The Tesla-class Pascal-100, Volta-100, and Ampere-100 continue to outperform our SEM-Accelerator by 4.3x, 6.41x, and 8.43x, respectively. It is interesting that the performance of the GPU kernel proposed in [40] seems to degrade for too high degrees, since it is only optimized for relevant polynomial degrees.

Finally, Fig. 2 provides an overview of the peak performance of all systems for polynomial order 7, 11, and 15 [7], including the calculated performance per unit watt (GFLOP/s/Watt). For $N = 15$, the SEM-accelerator provides higher performance than any of the general-purpose CPUs, as well as the NVIDIA K80, while being competitive with the RTX 2060 (211 vs. 244 GFLOP/s). For polynomial degree 11, only the Intel Xeon 6130 is faster than our SEM-accelerator, and at $N = 7$, only Marvell ThunderX2 is slower than our accelerator. The GPUs, in particular Pascal-100, Volta-100, and Ampere-100, rule supreme across all architectures for this type of application. Overall though, compared to the GPUs, the utilized bandwidth on the FPGA was higher as a percentage of theoretical bandwidth [40], if this continues to be the case for higher bandwidth speeds, this provides a case in favor for future FPGAs in memory bound applications. The results for our bandwidth measurements and other supplemental measurements can be found in an appendix[8].

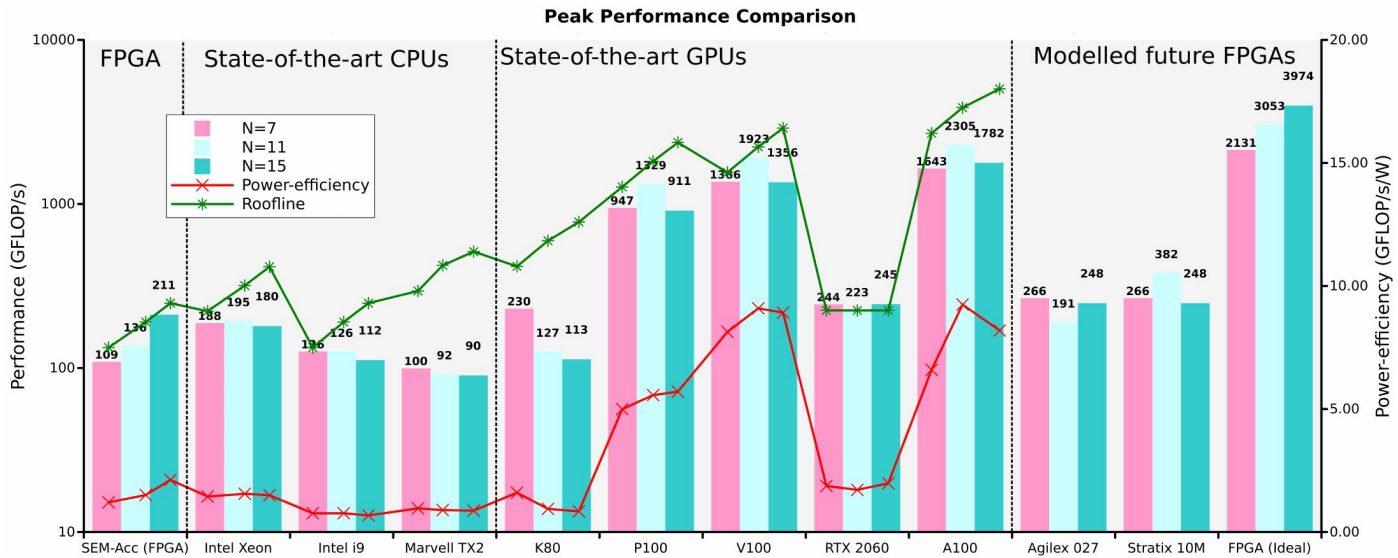

Fig. 2: Performance comparison at 4096 elements of the target application using different state-of-the-art architectures, showing measured performance (bars, left y-axis), power-efficiency (red-line, right y-axis), and theoretical roofline performance (green-line, left y-axis). Notice how our accelerator outperforms the general-purpose CPUs, while a hypothetical *ideal* FPGA could outperform even the latest Nvidia A-100.

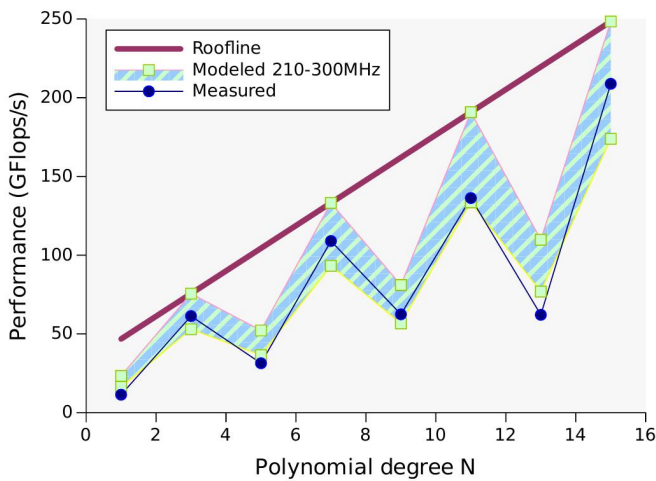

Fig. 3: Performance for our SEM-accelerator for 4096 elements compared to the theoretical roofline and the modeled performance at memory clock speed, 300 MHz, and 70% of the memory clock at 210MHz.

From a power efficiency perspective, our FPGA-based SEM-accelerator is more power-efficient than all the general-purpose CPUs, including the NVIDIA K80 (albeit not for $N = 7$), and rivals the GTX 2060 in power-efficiency for $N = 11$, and beating it for $N = 15$. The Tesla-class GPUs, including Pascal-100, Volta-100, and Ampere-100, have the highest power-efficiency, being up-to 2.69x, 4.44x, and 4.52x more power-efficient than our SEM-accelerator, respectively. Note, however, that the GPUs provide a lesser power efficiency benefit compared to their absolute performance; for example, the Pascal-100 is up-to 4.3x faster than our SEM-accelerator but only provides 2.69x the power-efficiency.

*To summarize:* **If peak performance and power-efficiency is a priority**, then the latest three generations of Tesla-class GPUs, without doubt, provide the highest observable performance as well as the highest power-efficiency. If, however, there are other constraints such as space, heat dissipation, or co-existence with other applications with other constraints (e.g., data-centers), the FPGAs are an excellent candidate to provide benefits over using general-purpose CPUs in both power- and performance, even for double-precision computations.

*D. Future and Hypothetical FPGAs: What would it take to beat the Ampere-100 using an FPGA?*

We end our result section by investigating where the FPGAs are heading through our performance model. Recall that our performance model (introduced in section IV) projects the performance of our SEM-accelerator given an FPGA with different resources (Logic, DSPs, BRAMs, and external bandwidth). This performance model allows us to *vary* the different parameters in order to not only predict performance using our current FPGA but also other FPGAs. Using our performance model and the experimental resource utilization we have on the Stratix 10, we project the performance of three devices. The first device is an Intel Agilex 027, which is the successor to Intel's Arria [9] family and a generation ahead of Stratix 10, which we couple with an external memory capable of delivering 153.6 GB/s (e.g., similar to what Marvell ThunderX2 has, see Table II). The second device is a recent variant of the Stratix 10 (10M [10]) device that is explicitly used for prototyping ASICs, which means that there is an ample amount of logic on the device (factor 3.6x larger than our current FPGA), has 5.7k DSP blocks (40 %fewer than our current), and we couple it with a 306 GB/s external memory (slightly higher than NVIDIA's K80, see Table II, but lower than RTX 2060). Our final device is a device that does not exist today; instead, we ask ourselves the question: how would the FPGA device look that would beat or be comparable to the Ampere-100 in performance for SEM-based computations? The device outputted by our performance model is a device with 6.2 million ALMs (factor 6x larger than our current), has 20k DSPs (4 times more than our current), and 12.9k BRAMs (only 10% more than our current), excluding the static partition, and is driven with an external memory supporting 1.2 TB/s (which is less than Ampere-100's 1.5 TB/s, see Table II). For all projections, we assume a mere 300 MHz clock frequency.

---

[7] These were chosen because these are where we were able to avoid arbitration on our SEM-accelerator, and because it is for $N$ in the range 7-11 that most SEM simulations are made [41].

[8] http://urn.kb.se/resolve?urn=urn:nbn:se:kth:diva-284225

These three devices performance is plotted in Fig. 2. We see that the upcoming Intel Agilex 027 is projected to be outperforming all CPUs and the K80 GPU, but still be far from capable of reaching NVIDIA Pascal-100 (and beyond) performance. The estimated peak performance for Intel Agilex 027 running our SEM-accelerator is 266, 191 and 248 GFLOP/s, and the device is logic-bound. The nature of the Agilex's performance is because of the constraint that the throughput must be a power of 2. Even if the device can support a throughput of, say 6 this is reduced down to 4, leading to lower performance for $N = 11$. The Agilex's perfomance illustrates clearly how, even if our SEM accelerator is bound by bandwidth for low degrees, the Agilex does simply not have enough logic to match the increased bandwidth as the polynomial degree increases. The Stratix 10M – a device created for ASIC prototyping – is projected to reach only slightly higher performance the Agilex, peaking at 382 GFlops/s at $N = 11$. Amusingly, unlike the Agilex, the execution on this device is actually DSP-bound, and if Intel should choose to create the same device, but with 8.7k DSPs (Only slightly more than the Agilex's) and increase the external bandwidth to 600 GB/s (on par with NVIDIA P100), then the modeled performance would be up to 1.06, 1.53, and 0.99 TFLOP/s respectively, which would be on par with or outperform the current implementation for the NVIDIA Pascal-100. Finally, our hypothetical FPGA – should it ever be incepted – would be able to actually outperform the NVIDIA Ampere-100 running the current GPU implementation of our kernel. The final performance for such hypothetical FPGA would, exactly like the A100, be memory bound, but also DSP/logic bound depending on the polynomial degree with a theoretical peak performance of 2.1, 3, 3.97 TFLOP/s, rivaling the roofline for the A100 based on its 1555 GB/s bandwidth.

*In summary:* **Can FPGAs ever come close to GPUs for SEM-based calculations?** Yes, they can, if the manufacturers allocate the resources differently than how they are allocated today. In particular, for these types of computations, there is a need to have a higher logic-to-DSP ratio, and, as calculated, the Stratix 10M device with more DSPs would actually rival the NVIDIA Volta-100 in performance. This, of course, assumes that the external memory of the FPGAs also increase, albeit could need less bandwidth than the GPUs to reach the same performance (as seen in our projection). Finally, there is always the opportunity for the manufacturers to specialize their DSP blocks to double-precision (similar to how Intel specialized DSPs blocks to single-precision), which would reduce the pressure on the logic and likely make the computation memory-bound, comparable to that of the GPUs.

## VI. RELATED WORK

There has been ample work focusing on Computational Fluid Dynamics (CFD) on FPGAs. Unfortunately, a majority of these focus on formulating the problem as a Stencil computation [43]–[45], (e.g., often solving the problem using the Lattice-Boltzmann Method). Stencil accelerations have previously been shown to map very well to FPGA-based systems [19], [46], in particular for IEEE-754 single-precision computations, and most studies limit themselves to study this form of CFD on FPGAs. Unlike LBM-type methods, our work focuses on high-order methods on unstructured meshes. Recently, Kenter et al. [47] was among the first to study the acceleration of the Discontinuous Galerkin (DG) method on FPGAs, reaching up to 164 and 103 GFLOP/s of performance for the volume and surface kernel, respectively, for single-precision computations. Nagy et al. [48] introduced a custom hardware unit for the acceleration of 2D Euler flow equations, reaching performance similar to an Intel Xeon E5620 on double-precision arithmetic. A similar, earlier solution by Sano et al. [49] leverages a custom 2D systolic array operating on single-precision to accelerate 2D Euler flow problems, reaching a peak performance of 11.5 GFLOP/s. Both of these reached performance levels that – for their time – was quite high.

## VII. CONCLUSION

In this work, we have described our implementation for an FPGA accelerator targeting high-fidelity SEM-based calculations in *double-precision* and modeled its performance from first principles. We showed that our accelerator implemented on state-of-the-art FPGAs could outperform (in both performance and power-efficiency) modern server-class CPUs (Intel Xeon and Marvell ThunderX2) and compete with consumer-grade Turing-class GPUs. We also positioned the performance against modern Tesla-class GPU accelerators, including the new Ampere-100 GPU, and show that FPGAs – particularly for double-precision computations – still have the potential for improvement. Using our performance-model, we conclude by projecting the performance of upcoming FPGAs, including a "hypothetical" FPGA, ultimately describing the characteristics of what a future FPGA should look like to outperform the Nvidia Ampere-100 on these types of calculations.

## ACKNOWLEDGMENT

Financial support from the SeRC Exascale Simulation Software Initiative (SESSI) is gratefully acknowledged. The computations were performed on resources provided by the Swedish

---

[9] https://www.intel.com/content/dam/www/programmable/us/en/pdfs/literature/pt/intel-agilex-f-series-product-table.pdf
[10] https://www.intel.com/content/dam/www/programmable/us/en/pdfs/literature/pt/stratix-10-product-table.pdf

National Infrastructure for Computing (SNIC) at High Performance Computing Center North (HPC2N). The authors gratefully acknowledge the funding of this project by computing time provided by the Paderborn Center for Parallel Computing (PC$^2$). We thank Hamid Reza Zohouri for helpful advice and support.

## References

[1] G. E. Moore, "Cramming more components onto integrated circuits," *Electronics Magazine*, vol. 38, no. 8, 04 1965.

[2] M. Horowitz et al., "Scaling, Power, and the Future of CMOS," in *IEEE InternationalElectron Devices Meeting, 2005. IEDM Technical Digest.* IEEE, 2005, pp. 7–pp.

[3] T. N. Theis and H.-S. P. Wong, "The End of Moore's Law: A New Beginning for Information Technology," *Computing in Science & Engineering*, vol. 19, no. 2, pp. 41–50, 2017.

[4] D. Kirk et al., "Nvidia CUDA software and GPU parallel computing architecture," in *ISMM*, vol. 7, 2007, pp. 103–104.

[5] H. Esmaeilzadeh, E. Blem, R. S. Amant, K. Sankaralingam, and D. Burger, "Dark Silicon and the End of Multicore Scaling," in *International symposium on computer architecture (ISCA)*. IEEE, 2011, pp. 365–376.

[6] J. S. Vetter, E. P. DeBenedictis, and T. M. Conte, "Architectures for the Post-Moore Era," *IEEE Micro*, vol. 37, no. 4, pp. 6–8, 2017.

[7] S. Matsuoka et al., "From FLOPS to BYTES: disruptive change in high-performance computing towards the post-moore era," in *Proc. of the ACM International Conference on Computing Frontiers*, 2016, pp. 274–281.

[8] I. Kuon, R. Tessier, and J. Rose, *FPGA Architecture: Survey and Challenges*. Now Publishers Inc, 2008.

[9] T. S. Czajkowski et al., "From OpenCL to high-performance hardware on FPGAs," in *22nd international conference on field programmable logic and applications (FPL)*. IEEE, 2012, pp. 531–534.

[10] J. Fifield, R. Keryell, H. Ratigner, H. Styles, and J. Wu, "Optimizing OpenCL applications on Xilinx FPGA," in *Proceedings of the 4th International Workshop on OpenCL*, 2016, pp. 1–2.

[11] J. Huthmann, L. Sommer, A. Podobas, A. Koch, and K. Sano, "OpenMP Device Offloading to FPGAs Using the Nymble Infrastructure," in *International Workshop on OpenMP*. Springer, 2020, pp. 265–279.

[12] A. Podobas and M. Brorsson, "Empowering OpenMP with Automatically Generated Hardware," in *International Conference on Embedded Computer Systems: Architectures, Modeling and Simulation (SAMOS)*. IEEE, 2016, pp. 245–252.

[13] A. Canis et al., "From Software to Accelerators with LegUp High-Level Synthesis," in *2013 International Conference on Compilers, Architecture and Synthesis for Embedded Systems (CASES)*. IEEE, 2013, pp. 1–9.

[14] S. Lee, J. Kim, and J. S. Vetter, "OpenACC to FPGA: A Framework for Directive-Based High-Performance Reconfigurable Computing," in *2016 IEEE International Parallel and Distributed Processing Symposium (IPDPS)*. IEEE, 2016, pp. 544–554.

[15] T. Becker, O. Mencer, S. Weston, and G. Gaydadjiev, "Maxeler Data-Flow in Computational Finance," in *FPGA Based Accelerators for Financial Applications*. Springer, 2015, pp. 243–266.

[16] H. R. Zohouri, N. Maruyama, A. Smith, M. Matsuda, and S. Matsuoka, "Evaluating and Optimizing OpenCL Kernels for high Performance Computing with FPGAs," in *SC'16: Proceedings of the International Conference for High Performance Computing, Networking, Storage and Analysis*. IEEE, 2016, pp. 409–420.

[17] A. Podobas, H. R. Zohouri, N. Maruyama, and S. Matsuoka, "Evaluating High-Level Design Strategies on FPGAs for High-Performance Computing," in *2017 27th International Conference on Field Programmable Logic and Applications (FPL)*. IEEE, 2017, pp. 1–4.

[18] C. Plessl, "Bringing FPGAs to HPC Production Systems and Codes," in *Fourth International Workshop on Heterogeneous High-performance Reconfigurable Computing, workshop at Supercomputing*, 2018.

[19] H. R. Zohouri, A. Podobas, and S. Matsuoka, "Combined Spatial and Temporal Blocking for High-Performance Stencil Computation on FPGAs using OpenCL," in *Proceedings of the 2018 ACM/SIGDA International Symposium on Field-Programmable Gate Arrays*, 2018, pp. 153–162.

[20] T. Kenter, J. Förstner, and C. Plessl, "Flexible FPGA design for FDTD using OpenCL," in *Proc. Int. Conf. on Field Programmable Logic and Applications (FPL)*. IEEE, 2017, pp. 1–7.

[21] L. C. Stewart, C. Pasoe, B. W. Sherman, M. Herbordt, and V. Sachdeva, "An OpenCL 3D FFT for Molecular Dynamics Simulations on Multiple FPGAs," *arXiv preprint arXiv:2009.12617*, 2020.

[22] C. Yang et al., "Fully integrated FPGA molecular dynamics simulations," in *Proceedings of the International Conference for High Performance Computing, Networking, Storage and Analysis*, 2019, pp. 1–31.

[23] P. Gorlani, T. Kenter, and C. Plessl, "OpenCL Implementation of Cannon's Matrix Multiplication Algorithm on Intel Stratix 10 FPGAs," in *2019 International Conference on Field-Programmable Technology (ICFPT)*. IEEE, 2019, pp. 99–107.

[24] Z. Jin and H. Finkel, "Exploring the Random Network of Hodgkin and Huxley Neurons with Exponential Synaptic Conductances on OpenCL FPGA Platform," in *2019 IEEE 27th Annual International Symposium on Field-Programmable Custom Computing Machines (FCCM)*. IEEE, 2019, pp. 316–316.

[25] A. Podobas and S. Matsuoka, "Designing and Accelerating Spiking NeuralNetworks using OpenCL for FPGAs," in *2017 International Conference on Field Programmable Technology (ICFPT)*. IEEE, 2017, pp. 255–258.

[26] Y. Kodama, T. Hanawa, T. Boku, and M. Sato, "PEACH2: An FPGA-based PCIe network device for Tightly Coupled Accelerators," *ACM SIGARCH Computer Architecture News*, vol. 42, no. 4, pp. 3–8, 2014.

[27] Y. Umuroglu et al., "FINN: A Framework for Fast, Scalable Binarized Neural Network Inference," in *Proceedings of the 2017 ACM/SIGDA International Symposium on Field-Programmable Gate Arrays*, 2017, pp. 65–74.

[28] C. Wang et al., "DLAU: A Scalable Deep Learning Accelerator Unit on FPGA," *IEEE Transactions on Computer-Aided Design of Integrated Circuits and Systems*, vol. 36, no. 3, pp. 513–517, 2016.

[29] J. D. Anderson and J. Wendt, *Computational Fluid Dynamics*. Springer, 1995, vol. 206.

[30] J. Shin et al., " Speeding Up Nek5000 with Autotuning and Specialization," in *Proceedings of the 24th ACM International Conference on Supercomputing*, 2010, pp. 253–262.

[31] J. W. L. Paul F. Fischer and S. G. Kerkemeier, "nek5000 Web page," 2008, http://nek5000.mcs.anl.gov.

[32] K. Świrydowicz et al., "Acceleration of tensor-product operations for high-order finite element methods," *The International Journal of High Performance Computing Applications*, vol. 33, no. 4, pp. 735–757, 2019.

[33] libCEED, "Development site," https://github.com/ceed/libceed, 2020.

[34] P. Fischer and K. Heisey, "NEKBONE: The Thermal Hydraulics mini-application," https://github.com/Nek5000/Nekbone, 2013.

[35] M. O. Deville, P. F. Fischer, and E. H. Mund, *High-Order Methods for Incompressible Fluid Flow*. Cambridge University Press, 2002.

[36] M. Butrashvily. CLFORTRAN. Accessed: Oct. 10, 2020. [Online]. Available: https://github.com/cass-support/clfortran

[37] J. Gong et al., "Nekbone performance on GPUs with OpenACC and CUDA fortran implementations," *J. Supercomput.*, vol. 72, 07 2016.

[38] H. R. Zohouri and S. Matsuoka, "The Memory Controller Wall: Benchmarking the Intel FPGA SDK for OpenCL Memory Interface," in *2019 IEEE/ACM International Workshop on Heterogeneous High-performance Reconfigurable Computing (H2RC)*, 2019, pp. 11–18.

[39] S. Williams, A. Waterman, and D. Patterson, "Roofline: An Insightful Visual Performance Model for Floating-Point Programs and Multicore Architectures," *Commun. ACM*, vol. 52, no. 4, pp. 65–76, 2009.

[40] M. Karp, N. Jansson, A. Podobas, P. Schlatter, and S. Markidis, "Optimization of Tensor-product Operations in Nekbone on GPUs," *arXiv preprint arXiv:2005.13425*, 2020.

[41] N. Offermans et al., "On the Strong Scaling of the Spectral Element Solver Nek5000 on Petascale Systems," in *Proc. EASC'16*, 2016, pp. 1–10.

[42] M. Meyer, T. Kenter, and C. Plessl, "Evaluating FPGA Accelerator Performance with a Parameterized OpenCL Adaptation of the HPCChallenge Benchmark Suite," *arXiv preprint arXiv:2004.11059*, 2020.

[43] C. Du, I. Firmansyah, and Y. Yamaguchi, "FPGA-Based Computational Fluid Dynamics Simulation Architecture via High-Level Synthesis Design Method," in *International Symposium on Applied Reconfigurable Computing*. Springer, 2020, pp. 232–246.

[44] K. Sano and S. Yamamoto, "FPGA-Based Scalable and Power-efficient Fluid Simulation using Floating-Point DSP Blocks," *IEEE Transactions on Parallel and Distributed Systems*, vol. 28, no. 10, pp. 2823–2837, 2017.

[45] H. M. Waidyasooriya, Y. Takei, S. Tatsumi, and M. Hariyama, "OpenCL-based FPGA-platform for Stencil Computation and its Optimization Methodology," *IEEE Transactions on Parallel and Distributed Systems*, vol. 28, no. 5, pp. 1390–1402, 2016.

[46] H. Fu, R. G. Clapp, O. Mencer, and O. Pell, "Accelerating 3D Convolution using Streaming Architectures on FPGAs," in *SEG Technical Program Expanded Abstracts 2009*. Society of Exploration Geophysicists, 2009, pp. 3035–3039.

[47] T. Kenter et al., "OpenCL-based FPGA Design to Accelerate the Nodal Discontinuous Galerkin Method for Unstructured Meshes," in *2018 IEEE 26th Annual International Symposium on Field-Programmable Custom Computing Machines (FCCM)*. IEEE, 2018, pp. 189–196.

[48] Z. Nagy et al., "FPGA Based Acceleration of Computational Fluid Flow Simulation on Unstructured Mesh Geometry," in *Int. Conference on Field Programmable Logic and Applications (FPL)*. IEEE, 2012, pp. 128–135.

[49] K. Sano, T. Iizuka, and S. Yamamoto, "Systolic Architecture for Computational Fluid Dynamics on FPGAs," in *15th Annual IEEE Symposium on Field-Programmable Custom Computing Machines (FCCM 2007)*. IEEE, 2007, pp. 107–116.